# Spin-torque driven magnetization dynamics in a nanocontact setup for low external fields: numerical simulation study

D.V. Berkov, N.L. Gorn

*Innovent Technology Development e.V.,
Pruessingstr. 27B, D-07745 Jena, Germany*

## Abstract

We present numerical simulation studies of the steady-state magnetization dynamics driven by a spin-polarized current in a point contact geometry for the case of a relatively large contact diameter ($D$ = 80 nm) and small external field ($H$ = 30 Oe). We show, that under these conditions the magnetization dynamics is qualitatively different from the dynamics observed for small contacts in large external fields. In particular, the 'bullet' mode with a homogeneous mode core, which was the dominating localized mode for small contacts, is not found here. Instead, all localized oscillation modes observed in simulations correspond to different motion kinds of vortex-antivortex (V-AV) pairs. These kinds include rotational and translational motion of pairs with the V-AV distance $d \sim D$ and creation/annihilation of much smaller (satellite) V-AV pairs. We also show that for the geometry studied here the Oersted field has a qualitative effect on the magnetization dynamics of a 'free' layer. This effect offers a possibility to control magnetization dynamics by a suitable electric contact setup, optimized to produce a desired Oersted field. Finally, we demonstrate that when the magnetization dynamics of the 'fixed' layer (induced only by the stray field interaction with the 'free' layer) is taken into account, the threshold current for the oscillation onset is drastically reduced and new types of localized modes appear. In conclusion, we show that our simulations reproduce semiquantitatively several important features of the magnetization dynamics in a point contact system for low external fields reported experimentally.

# I. INTRODUCTION

Magnetization dynamics induced in thin multilayer elements by a spin polarized current (SPC) is currently one of the most intensively studied topics in the solid state magnetism. After theoretical predictions [1, 2] and first experimental confirmations of this phenomenon [3] it was quickly realized that SPC-induced magnetization excitations and switching represent not only a very interesting phenomenon from the fundamental point of view (see review papers [4, 5, 6, 7], but is also are a very promising candidate for numerous device applications (for recent reviews see [8, 9, 10, 11]).

Among various experimental geometries used to study the spin torque induced magnetization dynamics, the so called point contact setup (where the current is injected into a multilayer element with lateral sizes in the mkm-region via a contact with the diameter $D_c \sim$ 10 - 100 nm) is one of the most interesting designs: It offers a large variety of magnetization oscillation modes, depending on the contact size, applied field strength and direction and magnetic materials used to compose a multilayer element [12, 13, 14, 15]. Moreover, one of the desired applications of the SPC-induced dynamics is the construction of *dc*-fed microwave generators, which should output enough power to be applicable in real technical devices. For this application the point contact setup is especially interesting: there exist - at least in principle - a possibility to synchronize magnetization oscillations induced by several point contacts attached to the same multilayer [16, 17], thus greatly increasing the output power due to the constructive wave interference.

For all these reasons magnetization dynamics in point contact devices has been the subject of an intensive research during the last few years. Already the first reliable experimental observation of magnetization oscillation in this geometry [12] posed an intriguing question about the nature of the observed oscillation mode. Namely, the measured oscillation frequency, being below the homogeneous FMR frequency for the magnetic layer studied in [12], could not correspond to the propagating wave mode predicted for such devices by Slonczewski [18]. Numerical simulations have shown that in such systems at least one *localized* mode type could exist [19] (in addition to the propagating wave), which was independently identified by the analytical theory [20] as a non-linear 'bullet'. Further theoretical and numerical studies have proven [21, 22] that this 'bullet' was indeed the mode observed in the pioneering paper [12].

Detailed numerical simulations [22, 23] have suggested, that a much more complicated localized modes, consisting of vortex-antivortex pairs, can exist in point contact devices. The frequency of these modes, according to our simulation predictions, should be much lower than for the 'bullet' mode (not to mention the propagating wave mode), and, should be also nearly current-independent [22]. Both features would make these modes very interesting from the point of view of technical application, because such modes would expand the frequency range of SPC-based microwave generators and offer a stability of the generated frequency with respect to current strength fluctuations. Further experimental studies have indeed shown, that magnetization oscillations observed in the nanocontact setup in very weak in-plane [14, 24] or out-of-plane fields [14, 25] can not correspond to the 'bullet' mode. Due to their very low frequency $f \sim$ 100 - 500 MHz (what is really low for the SPC-induced dynamics), the oscillations observed in Ref. [14, 25] were attributed to magnetization dynamics governed by the motion of a single vortex (for a strong out-of-plane field in [25]) or vortex-antivortex pairs (in very weak fields [24, 26]).

Especially interesting in this context are magnetization oscillations observed in point-contact devices with relatively large contact diameter (~ 100 nm) and in weak external fields. In such conditions, due to the absence of the stabilizing influence of the external field and relatively large area flooded by a spin polarized current, strongly inhomogenous magnetization excitations can occur. Indeed, a qualitatively new oscillation mode was found experimentally in [14], where for a point contact with the diameter ≈ 60 - 80 nm, attached to an extended $Co_{90}Fe_{10}/Cu/Ni_{80}Fe_{20}$ multilayer, microwave oscillations with extremely low frequency (down to ~ 100 MHz) and very weak frequency dependence on the current strength were observed. It was suggested that the observed dynamics can be explained by the generation and movement of a magnetic vortex, but no supporting simulations or analytical theory were reported in [14].

For all the reasons explained above, we have performed systematic numerical studies of the SPC-induced magnetization excitations for the case, when a point contact with a relatively large diameter is attached to an extended multilayer and the applied field is very weak (some preliminary results of this research have been briefly reported in the last part of our overview [26]). The paper is organized as follows. In Sec. II we describe in detail the simulation methodology, geometry and magnetic parameters of the simulated system. Sec. III contains the description of our main results: we start with the single layer system without the Oersted field (subsection III.A), proceed with the demonstration of the Oersted field effects (part III.B) and finish our presentation with the analysis of magnetodipolar interaction effects, when the 'hard' magnetic layer is included into the simulated system. Sec. IV contains the comparison of our results with experimental data and numerical simulations of other groups.

## II. SIMULATED SYSTEM AND SIMULATION METHODOLOGY

As mentioned in the Introduction, in this paper we intend to study the influence of various physical factors on the magnetization precession induced in the point contact geometry by a spin-polarized current (SPC) flowing perpendicular to the magnetic multilayer plane. In contrast to most previous numerical simulation studies [21, 22, 23, 25, 27], we focus our attention on the case where (*i*) the diameter of the point contact $D_c$ is relatively large and (*ii*) the external field $H_{ext}$ is small compared to the saturation magnetization of the material. As expected from general arguments and confirmed by recent experiments of the NIST group [14], for this case one can expect a qualitatively different magnetization dynamics as compared to systems with smaller point contacts and high external fields. In particular, both the larger value of the contact diameter and the smaller strength of the external field should allow for more complicated magnetization configurations, leading to an even richer set of non-linear localized modes than that reported in [21, 22, 25]. For this reason and keeping in mind at least a qualitative comparison with the experiments performed in [14], we have chosen the value $D_c = 80$ nm for the contact diameter and $H_{ext} = 30$ Oe for all result sets presented below.

This relatively large value of $D_c$ allowed us to choose larger lateral size of the discretization cell than in Ref. [22]: in the present study we use the mesh with the in-plane cell size 5 x 5 $nm^2$. In order to understand the influence of the interlayer interaction on the SPC-induced magnetization dynamics, we have studied both a *single* layer system and the *complete magnetic trilayer* consisting of materials as used in [14]. For the single layer system magnetic parameters corresponding to $Ni_{80}Fe_{20}$ (Permalloy) (saturation magnetization $M_S = 640$ G, exchange stiffness constant $A = 1 \cdot 10^{-6}$ erg/cm, negligible magnetocrystalline anisotropy) and the layer thickness $h_{free} = 5$ nm were used. For the trilayer system we have adopted the parameters of the 'fixed' (hard) layer as for $Co_{90}Fe_{10}$, namely $M_S = 1500$ G [28], $A = 2 \cdot 10^{-6}$ erg/cm (various sources give the values of the CoFe exchange stiffness in the range $A = (1 - 3) \cdot 10^{-6}$ erg/cm), the layer thickness $h_{fix} = 20$ nm and the spacer thickness (distance between two magnetic layers) $h_{Cu} = 4$ nm [14]. Both magnetic layers were not discretized further into sublayers; it was checked that such a discretization, leading to a large increase of the computational time, did not significantly affect the results.

In order to avoid the artificial influence of the system borders, we have used periodic boundary conditions (PBC). To suppress the spin wave propagation between different PBC-replica, we employ the damping parameter which increased towards the simulated area borders as described in [22, 23]. In the present study the spatially dependent damping coefficient was chosen in the form $\lambda(r) = \lambda_0 + \Delta\lambda \cdot [1 + \tanh((r - R_0)/R_{dec})]$, where $r$ denotes the distance to the point contact center, $\lambda_0 = 0.02$ is the 'native' material damping constant, $R_0 = 1000$ nm, $R_{dec} = 100$ nm and the damping increase parameter is $\Delta\lambda = 1.0$. It has turned out that the usage of this damping profile and the lateral size of the simulated area $L \times L = 2000 \times 2000$ nm$^2$ is sufficient to fully suppress the above mentioned artificial spin wave interference arising due to PBC.

We have also studied the influence of the Oersted field of the *dc*-current flowing within the contact area (see Sec. III.B). Unfortunately, the electric current distribution in the experimental setup is not known exactly, so we could not compute the corresponding field directly from this distribution. For this reason, we had to adopt another strategy to study the Oersted field effect, which is explained at the beginning of Sec. III.B.

The magnetization dynamics itself was simulated using our commercially available micromagnetic package (see [29] for implementation details) with the extensions allowing to use (*i*) the site-dependent damping constant as explained above and (*ii*) the site-dependent current density in order to mimic the current flowing through the point contact area only. Thermal fluctuations were neglected ($T = 0$). Spin torque acting on the *free layer only* was included by adding the Slonczewski torque term $\Gamma_{st} = a_J [\mathbf{M} \times [\mathbf{M} \times \mathbf{p}]]$ to the 'normal' Landau-Lifshitz-Gilbert (LLG) equation. The amplitude of this torque $a_J$ is proportional to current strength $I$ and its spin polarization degree $P$ and depends also on the magnetic layer thickness $h$, the contact area $S_c = \pi R_c^2$, and magnetization $M_S$ (see, e.g., [30]):

$$a_J = \frac{\hbar \cdot I \cdot P}{2|e| M_S^2 \cdot h \cdot S_c} \tag{1}$$

The spin polarization direction of electrons **p** in the *dc*-current flowing through the contact area was chosen to be *opposite* to the applied field direction $\mathbf{H}_{ext}$ for a *single* layer system and *opposite* to the *local instantaneous magnetization direction* of the fixed layer for a *trilayer* system. The reason for this choice is the following: the magnetization of a free layer in real experiments is supposed to be excited by spin-polarized electrons *reflected* from the fixed magnetic layer (of a trilayer system) towards the free one.

### III. RESULTS AND DISCUSSION

#### A. Magnetization dynamics of the single-layer system in the absence of the Oersted field

In order to understand the influence of various physical factors on the magnetization dynamics separately, we proceed in a usual way, 'switching on' these factors in turn, thus isolating corresponding effects. Hence we start our study with the simplest system including the free magnetic layer only and neglecting the Oersted field of the spin-polarized current.

Oscillation power spectra for this system (Py layer with the thickness $h_{Py} = 5$ nm, subject to a spin-polarized current flowing through the point contact with the diameter $D_c = 80$ nm, placed into the in-plane external field $H_{ext} = 30$ Oe) are shown in Fig. 1. This figure displays the oscillation spectra of the $m_z$ magnetization component, whereby the *x*-axis is chosen along the external field direction, and the *z*-axis is directed in the film plane perpendicular to $\mathbf{H}_{ext}$.

#### *1. Propagating wave mode*

The first mode observed after the oscillation onset is the 'normal' propagating wave mode $W_1$ (Slonczewski mode) predicted in [18]; the index in the notation $W_1$ is necessary to distinguish this mode in the single-layer system from analogous propagating wave modes in other systems considered below. It can be seen that this mode exists in the relatively broad current range ($I_W \approx 14 - 20$ mA) and its frequency decreases continuously with growing current (we point out for clarity that the discrete character of the $f(I)$-dependence for this mode seen in Fig. 1 is an image artifact arising due to the discrete set of currents used in simulations and rapid decrease of the frequency with current).

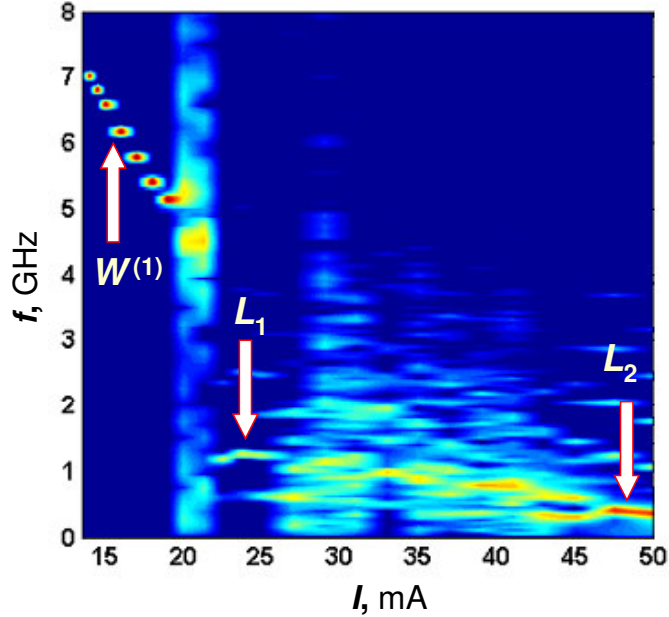

Fig. 1. Spectral power of the $m_z$-component (in-plane component perpendicular to the external field) as the function of the current strength in the single layer system without the Oersted field. Arrows indicate the position of modes analyzed below in more detail (see Fig. 2, 3, 4) The discrete character of the $f(I)$ dependence of the $W_1$ mode is an image artifact due to the discrete set of simulated currents and rapid frequency decrease with current for this mode.

The threshold current value for the oscillations onset is $I_{th} \approx 14$ *mA* and the corresponding threshold oscillation frequency is $f_{th} \approx 7.05$ GHz. It is instructive to compare these values to the analytical calculations for the Slonczewski mode [18, 20]. We remind that these calculations predict the threshold current

$$\sigma I_{th} \approx \frac{1.86 \cdot D(H_0)}{R_c^2} + \Gamma(H_0) \qquad (2)$$

where the factor $\sigma = g\mu_B P/(2|e|M_S \cdot h \cdot S_c)$ [20] is related to our spin torque amplitude $a_J$ as $\sigma I = \gamma M_S \cdot a_J$. The first term in (2) describes the (dominant) energy loss due to spin-wave emission by the point contact area and is proportional to the spin wave dispersion $D(H_0)$. For the field-in-plane geometry this dispersion reads

$$D(H_0) = \frac{2\gamma A}{M_S} \cdot \frac{H_0 + 2\pi M_S}{\sqrt{H_0(H_0 + 4\pi M_S)}} \qquad (3)$$

The energy losses within the point contact area due to the Gilbert damping are given by the second term in (2): $\Gamma(H_0) = \gamma \lambda \cdot (H_0 + 2\pi M_S)$; for the situation studied here these damping losses are much smaller than those due to first term.

For the system geometry ($h = 5$ nm, $R_c = 40$ nm), material parameters ($M_S = 640$ G, $A = 1 \cdot 10^{-6}$ erg/cm, $\lambda = 0.02$), current polarization $P = 0.4$ and external field $H_0 = 30$ Oe used in our simulations, the threshold current value calculated from this analytical theory is $I_{th}^{an} \approx 37$ mA. This result is about 2.5 times larger than the simulated value ($I_{th}^{sim} \approx 14$ mA).

Analytical prediction for the oscillation frequency [31] $\omega_{th}^{an} = \omega_0 + Dk_0^2$ contains the homogeneous FMR frequency, which for the in-plane-field is $\omega_0 = \sqrt{H_0(H_0 + 4\pi M_S)}$ and the wave vector of the excited circular spin wave, which was found to be [18, 20] $k_0 = 1.2/R_c$. In contrast to the large discrepancy for the threshold currents, this analytical result leads to the oscillation frequency $f_{th}^{an} = \omega_{th}^{an}/2\pi \approx 7.85$ GHz, the value only slightly above the frequency $f_{th}^{sim} = 7.05$ GHz observed numerically.

A possible reason for the large disagreement between the analytically predicted and numerically simulated threshold currents could be the approximations made by the derivation of Eq. (2) for $I_{th}^{an}$: first, it was obtained for the *perpendicularly* magnetized point contact, and second, it was assumed that the group velocity of emitted spin waves is *isotropic* with respect to the propagation direction. The partial adjustment of the Eq. (2) to our case of an in-plane magnetized contact could be achieved by using the expression (3) for the spin wave dispersion in case of an *in-plane* magnetization. However, the dependence of the spin wave velocity on the propagation direction could not be taken into account. We note that in our case the point contact diameter is relatively large, so that the wave vectors of the emitted spin waves $k \sim 1/R_c$ are relatively small. Hence the wave group velocity substantially depends on the angle between the wave propagation direction and external field (see gray-scale maps in Fig. 2). This high anisotropy of the group velocity could lead to significant discrepancies between analytical calculations and numerical data.

At the same time the expression for threshold frequency $\omega_{th}^{an} = \omega_0 + Dk_0^2$ uses - besides the in-plane spin wave dispersion factor $D$ - only the wave vector $k_0 \sim 1/R_c$, which exact value relies mainly on the geometrical consideration (circular shape of the current-flooded area). Hence the analytical prediction for the oscillation frequency should be more reliable, leading to a much better agreement between analytical theory and numerical simulations, as found above. These our arguments are supported, in particular, by analogous comparisons for the nanocontact with a smaller radius ($R_c$ = 20 nm) in [21, 22], where the anisotropy of the group velocity was much lower due to larger spin wave vectors. The agreement between analytically obtained and numerically simulated threshold currents for this case was, indeed, decisively better (see [21, 22] for details).

With increasing current the propagating wave mode demonstrates the strong downward frequency shift due to the growing oscillation amplitude. The frequency decreases from its initial value $f_{th} \approx 7$ GHz at $I_{th} \approx 14$ mA to $f \approx 5$ GHz reached for the current $I \approx 20$ mA, where the transition to localized modes occurs (see below). This frequency decrease with the growing current is nearly linear (we remind, that the jumps on the oscillation power plot in Fig. 1 are solely due to the discrete set of the current values used in simulations). This frequency decrease with increasing current due to the growing oscillation amplitude is a non-linear effect which is well understood theoretically [20, 32] and hence will not be further discussed here.

Concluding this discussion of the propagating wave mode, we would like to emphasize the modulation of the main wave profile by a 'secondary' wave with the vector $k \approx 2k_0$ approximately twice as large as the 'main' wave vector $k_0$, which can be clearly seen at the gray-scale maps displayed in Fig. 2. This modulation is due to the another non-linear effect arising due to the conservation of the local moment magnitude $M_x^2 + M_y^2 + M_z^2 = M_S^2$. This condition leads to the contribution of the spin wave with the frequency corresponding to $M_x$-oscillations to the wave pattern of the $M_z$-projection. The effect becomes more pronounced with the increase of the magnetization oscillation amplitude; the picture shown in Fig. 2 corresponds to $I$ = 16 mA, where the oscillation amplitude of the $M_z$-projection under the point contact is close to its maximal value $m_z^{max} = M_z^{max}/M_S \approx 1$.

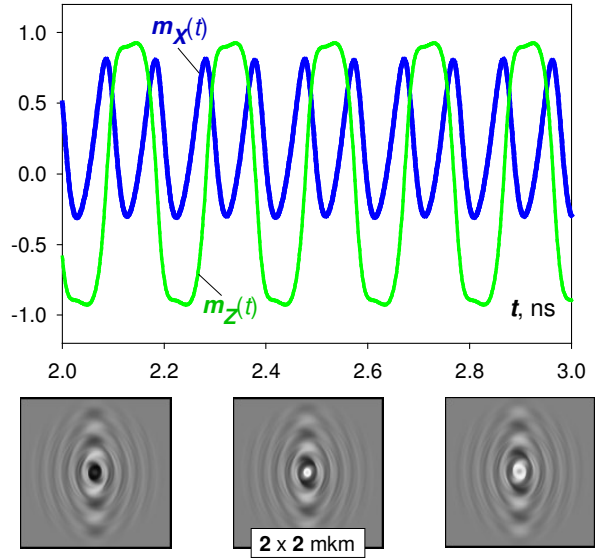

Fig. 2. Magnetization time dependencies and snapshots of the magnetization configurations as gray-scale images of the $m_z$-component for the propagating (Slonczewski) mode $W_1$ in the single layer system when the Oersted field is neglected.

## *2. Localized modes*

When the current strength exceeds the next critical value $I_{th}^{loc} \approx 20$ mA, a large frequency jump down to $f \approx 1.2$ GHz occurs. This frequency is well below the homogeneous FMR frequency $f_0 = (\gamma/2\pi)\sqrt{H_0(H_0 + 4\pi M_S)} \approx 1.4$ GHz for the film studied here, so that magnetization oscillations after the jump (above $I_{th}^{loc}$) should correspond to a localized mode. The spatiotemporal analysis of magnetization configurations reveals that all the modes occurring for $I > I_{th}^{loc}$ are indeed localized. Dynamical processes responsible for observed magnetization oscillations (see figures below) are qualitatively different for various modes. In many cases these processes are also highly irregular, so we describe and discuss below only those localized modes which are generated by relatively simple magnetization dynamics.

Magnetization oscillations for the first such mode, which appears after the transition from the propagating wave mode $W_1$ to localized oscillations, are shown in Fig. 3. First of all we point out, that this first localized mode ($L_1$-mode) has a completely different nature compared to the spin-wave 'bullet' observed in systems with the relatively small point contact diameter ($R_c$ = 20 nm) [20, 21, 22, 23]. We remind, that this localized 'bullet' mode has a relatively homogenous core magnetization structure, whereby the magnetization oscillation amplitude decreases exponentially with the distance from the contact center. In contrast to the 'bullet' mode, in our system the magnetization structure of the 1st localized mode is highly inhomogeneous: magnetization oscillations are caused by an appearance and rotation of a vortex-antivortex pair. This qualitative difference between the two systems studied here and in [20, 21, 22] is due to the small contact diameter $R_c$ = 20 nm and the large external field $H_{ext}$ = 2 kOe used in the paers cited above: both these factors strongly favor a homogeneous magnetization configuration of the localized mode core found in [20, 21, 22, 23]. In our case, where the point contact diameter is twice as large and the external field is nearly absent ($H_{ext}$ = 30 Oe), a formation of more complicated excitations - vortex-antivortex (V-AV) pairs - becomes possible.

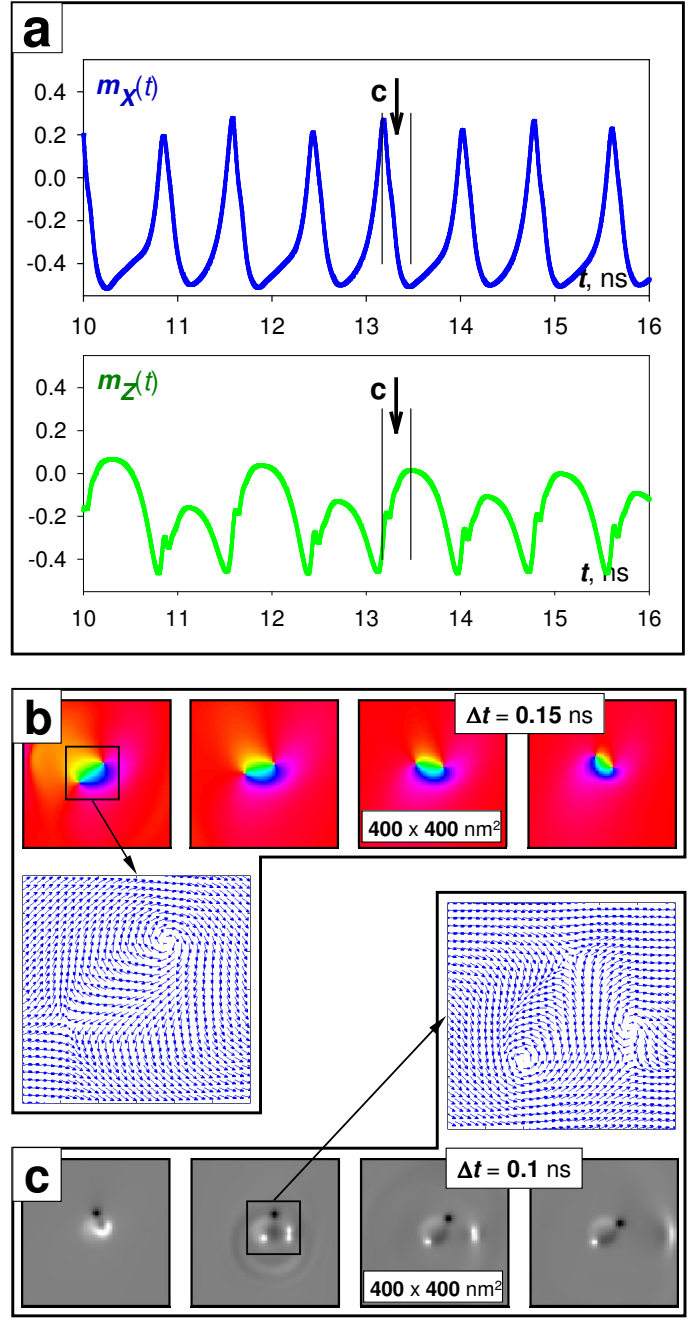

Fig. 3. Magnetization time dependencies (a) and snapshots of the magnetization configurations ((b) and (c)) for the first localized mode $L_1$ in the single layer system for $\mathbf{H}_{Oe} = 0$. Panel (b) shows the rotation process of the main vortex-antivortex pair (images of the in-plane magnetization orientations) together with the arrow plot of a typical V-AV configuration. Panel (c) illustrates the creation, propagation and decay of a satellite V-AV pair (gray-scale images of the out-of-plane magnetization orientations $m_y$).

As it can be seen from magnetization maps and arrow plots presented in Fig. 3, two major processes contribute to the magnetization dynamics of the $L_1$-mode in our system: (*i*) rotation of a vortex-antivortex pair with a relatively large V-AV distance (Fig. 3b) and (*ii*) generation and subsequent translational motion of a much smaller V-AV (Fig. 3c).

To start the analysis of these two processes, we first point out, that due to the periodic boundary conditions applied to the system and the homogeneous starting magnetization state, magnetic excitations with non-zero vorticity should appear in pairs in order to guarantee the conservation of the total topological charge [33, 34]. This situation is qualitatively different from those observed for finite small nanoelements (like nanodisks studied numerically in [25]), where due to the open boundaries and the out-of-plane external field the formation of a *single* vortex was possible.

Both rotational and translational motions of a V-AV pair mentioned above have been studied by Komineas and Papanicolaou [34, 35]. They have shown that a V-AV pair should *rotate* when vortex and antivortex have *opposite* polarities and undergo a *translational* motion when the polarities of a vortex and antivortex are the *same*. Results of our simulations agree with this analytical statement: as it can be seen from the gray-scale maps in Fig. 3c, polarities of vortex and antivortex are *opposite* for the large *rotating* V-AV pair, but their polarities *coincide* within the small *translationally* moving V-AV pair.

The next step is obviously the quantitative comparison of the rotation frequency and the translational velocity of a V-AV pair predicted in [34, 35] with our simulation results. In [34] it was shown, that a V-AV pair with vortex and antivortex having opposite polarities possesses a zero linear momentum (so that the pair center does not move), but a non-zero *angular* momentum **L**, so that such a pair should rotate (**L** increases as $\sim d^2$ with the V-AV separation distance $d$). The angular velocity of this rotation $\omega$ can be determined from the condition, that for this velocity the extended energy functional $F = E - \omega \mathbf{L}$ in the rotating coordinate system should possess a stationary point. The total magnetic energy of a V-AV pair in this functional $E = E_{ex} + E_{an}$ consists - in the approximations used in [34] (no external field, large V-AV separation) - from the sum of exchange $E_{ex}$ and the anisotropy $E_{an}$ energies of the vortex and antivortex. In the notation of Komineas $E_{an}$ includes also the demagnetizing energy in the thin-film geometry.

The form of the extended energy functional $F$ together with the scaling arguments similar to those used in the nonlinear dynamics of continuous media enables the determination of the rotation frequency $f$ of the V-AV dipole. Taking into account that the explanation given in the PRL-paper [34] was necessarily very brief and that scaling arguments mentioned above are not commonly familiar to the micromagnetic community, we present the basic line of these arguments here in order to make our paper self-contained. The main idea is, that the stationary point of the extended energy functional $F = E - \omega \mathbf{L}$ should ‚survive' the rescaling of spatial variables, which by itself does not change the physics of the system. In 2D systems (thin film limit) the exchange energy $E_{ex}$ is invariant with respect to such a rescaling, because $E_{ex}$ is a 2D integral over a *square* of the magnetization *gradient*. In contrast, $E_{an}$ is not invariant with respect to this transformation, being in the simplest case a 2D integral over the 2$^{nd}$ power of the magnetization components themselves. Hence the energy functional $F = E - \omega \mathbf{L} = E_{ex} + E_{an} - \omega \mathbf{L}$ may have a stationary point which is stable with respect to rescaling only when $E_{an} - \omega \mathbf{L} = 0$. Using this relation together with the expressions for the vortex anisotropy energy $E_{an}$ and the dependence of the angular momentum **L** on the V-AV separation $d$, Komineas [34] derived the rotation frequency for the V-AV pair, which in non-reduced units reads

$$f = 2 \cdot \frac{2}{\pi} \cdot \frac{\gamma A}{M_S d^2} \qquad (4)$$

Here $\gamma$ is the gyromagnetic ratio, $A$ - exchange stiffness and $M_S$ - saturation magnetization of the material. An additional factor of 2 (compared to the Eq. (11) in [34]) takes into account the magnetostatic interaction between vortex and antivortex, as discussed in the concluding part of [34].

Substituting into Eq. (4) magnetic parameter values used in our simulations ($M_S = 640$ G, $A = 1 \cdot 10^{-6}$ erg/cm) and the average value of the vortex-antivortex distance $d \approx 75$ nm determined from the simulated magnetization configuration like those shown in Fig. 3, we obtain the analytical value of the rotation frequency $f_{rot}^{an} \approx 0.62$ GHz.

The rotation period of the V-AV pair determined from the inspection of the simulated magnetization configurations turns out to be $T_{rot} \approx 1.60$ ns, which results in the simulated value of the rotation frequency $f_{rot}^{sim} \approx 0.62$ GHz. Such an excellent coincidence between analytically computed and simulated values of *f* is somewhat unexpected for the case studied here: neither the deformation of vortex and antivortex structures due to the V-AV interaction, nor the changing of the V-AV separation during the simulated pair rotation due to the presence of an external field, nor the process of the formation and emission of the 2$^{nd}$ V-AV pair (see below) are included into the analytical theory. So such a good agreement between simulations and analytical theory means, that (*i*) the approximation of a large V-AV separation adopted in [34] works in our case fairly well, (*ii*) the external field is weak enough not to disturb the free V-AV pair rotation and (*iii*) the process of a generation and emission of the satellite V-AV pair is very fast compared to the rotation period of the main V-AV pair.

Now we turn our attention to the second type of the V-AV pair dynamics - translational motion, also observed for the 1$^{st}$ localized mode $L_1$. Fig. 3c demonstrates that during the rotation of the main (large) V-AV pair another (much smaller) V-AV pair consisting of a vortex and antivortex with *the same* polarities is generated. Such a composite object has zero angular momentum (so that such a pair does not rotate) [35], but a *non-zero* linear momentum **P**, which results in a *translational* motion of this V-AV pair (the so called Kelvin motion, well known from the fluid dynamics). This translational motion can be clearly seen in Fig. 3c and its velocity can be compared to the analytical result from [35].

Analytical estimation made in [35] is based for the translational V-AV motion on the extended energy functional $F = E - v\mathbf{P}$ in a translationally moving coordinate system, where *v* is the linear velocity of the V-AV pair. The same scaling arguments concerning the stability of the stationary point of this functional with respect to the rescaling of spatial variables apply to this case also. Together with the expression of the anisotropy energy for the V-AV pair these arguments lead to the following estimation for the pair velocity in the limit of large V-AV separations:

$$v \sim 2 \cdot \frac{2\gamma A}{M_S d} \quad (5)$$

(here an additional factor of 2 results from the same V-AV interaction as explained in the text after the Eq. (4)). Substituting the same material parameter values and the V-AV separation *d* $\approx 25$ nm determined from the simulated magnetization configuration shown in Fig. 3c into the Eq. (5), we obtain the analytical estimation $v^{an} \sim 4.4 \cdot 10^4$ cm/s. The pair velocity for this system, measured from simulated magnetization configurations, is $v^{sim} \approx 6.3 \cdot 10^4$ cm/s. The significant difference between analytically calculated and simulated values of this V-AV pair is most probably due to the fact that for such closely placed vortex and antivortex, the deformation of their structures due to their mutual interaction plays a noticeable role (the V-AV separation is only about several exchange lengths $l_{ex} = (A/2\pi M_S^2) \approx 6.23$ nm).

An important circumstance is that this V-AV pair is gradually destroyed during its translational motion due to the presence of the finite energy dissipation, which could not be taken into account by the analytical theory [35]. This gradual decay is in contrast to the steady-state rotation of the large V-AV pair with the opposite V-AV polarities, which is due to the constant energy supply via the spin-polarized current. We also note that the formation and emission of this small V-AV pair is an important mechanism of the magnetic energy irradiation out of the point contact area.

When the current is increased further ($I > 26$ mA, see Fig. 1), a variety of more complicated and partially irregular localized modes appear. The overall trend is the increase of the number of V-AV pairs generated and annihilated per unit time. We postpone the detailed discussion of the intriguing magnetization dynamics in this current region to future publications and discuss only the regular localized mode $L_2$, observed for high current values (40-50 $mA$).

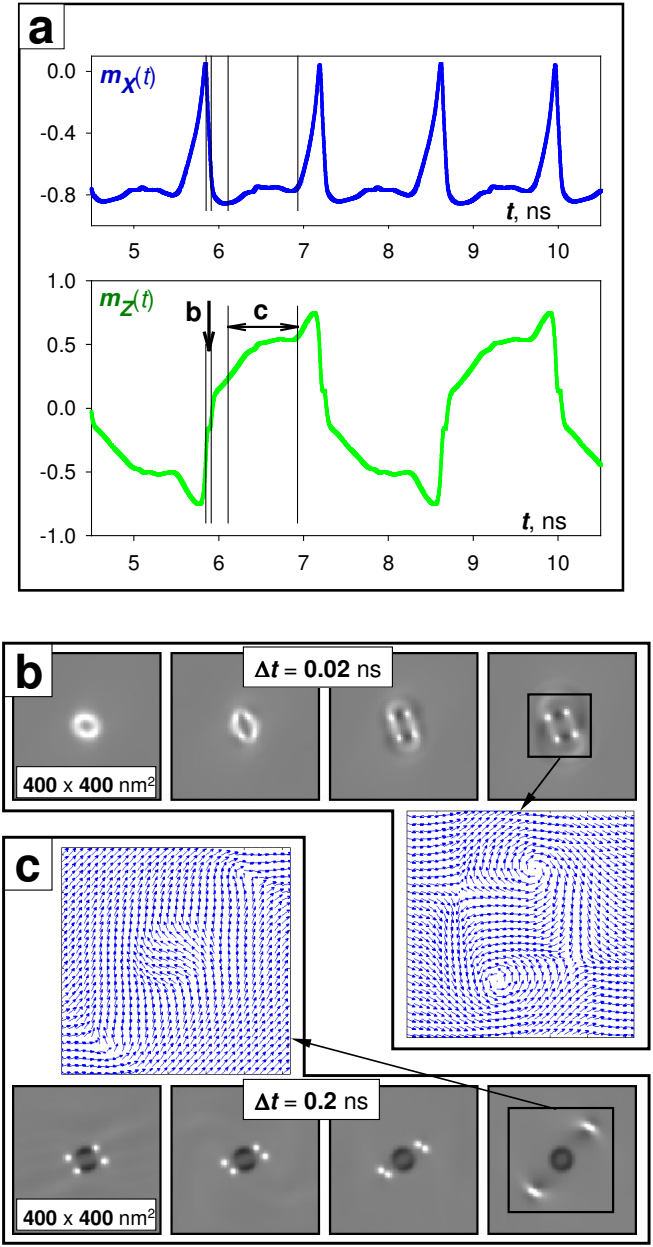

Fig. 4. Magnetization time dependencies (a) and gray-scale snapshots ((b) and (c)) of the out-of-plane magnetization component $m_y$ for the second localized mode $L_2$ (single layer, $\mathbf{H}_{Oe} = 0$). Panel (b) shows the creation process of two V-AV pairs and an in-plane arrow plot of a typical V-AV quadrupole configuration. Panel (c) displays the propagation and decay of the two V-AV pairs. Time intervals corresponding to the image rows of (b) and (c) are marked with vertical lines on $m_z(t)$- plot in the panel (a).

Such a strong current leads to the periodical creation/annihilation of a vortex-antivortex *quadrupole*, consisting of two vortices and two antivortices all having the same polarity (Fig. 4), and located symmetrically with respect to the point contact center. The formation of this V-AV quadrupole starts with the appearance of a ring-shaped magnetization structure (see the 1$^{st}$ gray-scale map in Fig. 4b), which evolves very fast into two V-AV pairs which form a nearly symmetrical V-AV quadrupole. From the point of view of non-linear excitation dynamics such a quadrupole represents the next possible stable excitation kind (after a single V-AV pairs) for systems where the total topological charge is initially zero and should be conserved. The whole process of the quadrupole formation shown in detail in Fig. 4b is very fast, taking only ~ 50 ps.

The quadrupole built this way does not rotate (because, as mentioned earlier, all its vortices and antivortices have the same polarity), so it can disappear only relatively slowly via the symmetry breaking with respect to the V-AV distances and Kelvin motion of the V-AV pairs arising after this symmetry breaking. The lifetime of the V-AV quadrupole limited by this process is ~ 1 ns as it can be seen from the magnetization time dependencies (Fig. 4a) and gray scale maps of the out-of-plane magnetization projections (Fig. 4c). The two solitons, each representing one V-AV pair, propagate in opposite directions and decay due to the 'normal' energy damping. These solitons are responsible for the energy emission out of the point contact area for this localized mode.

We note once more, that in all cases considered above the oscillation frequency of the $m_x$-component is two times larger than that of the $m_z$-component, as it can be clearly seen from Fig. 2a, 3a and 4a. This means, that by experimental observations of these modes using the GMR-effect, the strong second harmonics should be present not only due to the non-sinusoidal character of $m_z$-oscillations, but also due to the mixing of signals from $m_x$- and $m_z$-components. This mixing unavoidably happens when the magnetization of the fixed layer deviates (even slightly) from the external field direction, e.g., due to the influence of a random magnetocrystalline anisotropy of this fixed layer. As mentioned above, for $Co_{90}Fe_{10}$ often used in such experiments [12, 14], the cubic anisotropy constant $K_{cub} = - 5.6·10^5$ erg/cm$^3$ is not small [28], so that even in nanocrystalline materials noticeable deviations of the magnetization from its average direction (along the external field) can be expected.

### B. Magnetization dynamics for the single-layer system: Influence of the Oersted field

In this subsection we discuss the influence of the Oersted field, induced by the spin-polarized *dc*-current, on the magnetization dynamics. When a quantitative comparison with real experiments made on relatively large point contacts and in small external fields is aimed, one cannot neglect the influence of the Oersted field, because for this situation it is not small compared to the external field. Hence the corresponding influence can crucially change the magnetization dynamics, especially taking into account that the Oersted field of a thin wire is strongly inhomogeneous.

Unfortunately, it is not possible to compute the Oersted field quantitatively, because neither the experimental geometrical setup nor the current distribution in the contact leads are known exactly. For this reason we have adopted the following strategy in order to study the Oersted field effect. We are interested in the Oersted field acting on the free layer magnetization within the area directly under the point contact and in its vicinity. To compute the Oersted field for this region, we make use of the two circumstances. First, for a standard symmetrical lead setup, the Oersted field created by the current flowing within a particular lead far away from the point contact, is mostly cancelled by the corresponding field of the symmetrically located lead, where the current flows in the opposite direction. Second, in the lead region which is attached to the contact, the current flows from the lead into the point contact wire, which diameter is much smaller than the lateral lead size. So in a good qualitative approximation, the current flow within this lead region is similar to the water stream flowing from the large basin into a narrow drain channel. Hence for any particular current flow line there exist a symmetrically located flow line with the opposite current direction, so that the overall Oersted field of the lead area adjacent to the contact is also largely cancelled out.

The above arguments, which have been presented and quantitatively elaborated by Miltat [36], lead to the conclusion that the major contribution to the Oersted field under the point contact is due to the current flowing in the point contact itself. Still, there exist a possibility to change the Oersted field varying the height of the contact wire, so, in order to study the Oersted field effect, we proceed in the following way. First we calculate the field $\mathbf{H}_{inf}(I, r)$ generated by a current *I* in an *infinitely long* conductor with the diameter equal to that of the point contact (*r* denotes the distance from the contact center). Then, taking into account that the real contact wire has a finite length, we assume that the actual Oersted field has the same circular symmetry, but is weaker than $\mathbf{H}_{inf}(r)$. To take account of this weakening, we introduce the weakening coefficient $\kappa$ ($0 \leq \kappa \leq 1$), compute the Oersted field as $\mathbf{H}_{Oe}(r) = \kappa \cdot \mathbf{H}_{inf}(r)$ and study the magnetization dynamics for several values of this coefficient $\kappa$.

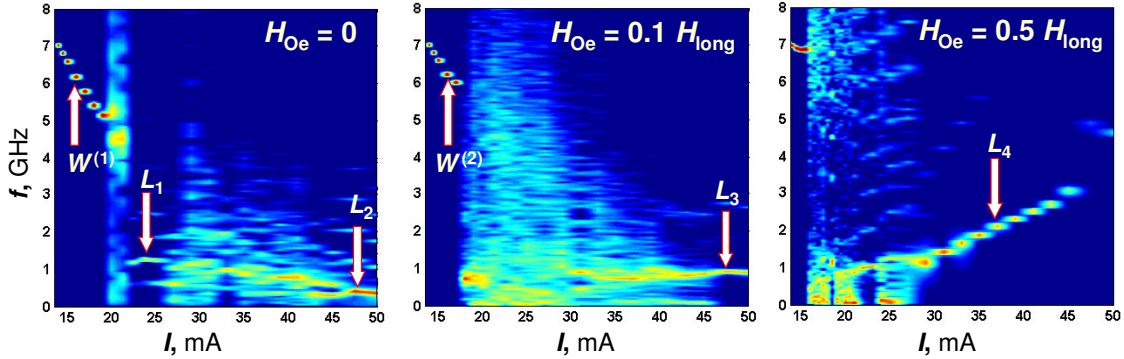

Fig. 5. Spectral power of the $m_z$-component vs current strength in the single-layer system for different strengths of the Oersted field computed as explained in the text. Arrows show the positions of modes analyzed in detail below. The discrete character of the $f(I)$ dependence for the $W_1$, $W_2$ and $L_4$ modes is an image artifact (see caption to Fig. 1).

We have found several qualitative effects of the Oersted field. To explain these effects in detail, we concentrate ourselves on two 'limiting' cases: relatively small ($\kappa = 0.1$) and large Oersted fields ($\kappa = 0.5$). Corresponding oscillation power maps in standard current-frequency coordinates are shown in Fig. 5 (middle and right panel) in comparison with the case when $\mathbf{H}_{Oe}$ is absent (left panel, the same as in Fig. 1).

These oscillation power plots allow to identify the following qualitative effects of the Oersted field on the magnetization dynamics:

(i) 'On average', the oscillation spectra become more regular, especially for large Oersted fields (see Fig. 5c, $\kappa = 0.5$), where well defined spectra with sharp peaks for nearly all current values are observed. This effect is due to the stabilizing influence of the Oersted field, which - at least for large $\kappa$-values - is strong enough to guarantee the existence of well defined modes, preventing the system from sliding into a quasichaotic behavior.

(ii) The current region, where the propagating wave mode exists, narrows with increasing $\kappa$. This contraction can be explained by a strongly non-homogeneous character of the Oersted field in the geometry under study, which induce a non-homogeneous and asymmetrical equilibrium magnetization configuration (in the absence of SPC-induced oscillations). Such a configuration obviously suppresses the oscillation mode represented by the wave propagating circularly out of the point contact area.

(iii). Localized modes containing V-AV pairs also behaves themselves qualitatively different in presence of the Oersted field, because this circular field acts differently on a vortex and an antivortex, thus disturbing a free rotation of a V-AV pair (see discussion below).

To study the magnetization dynamics in presence of the Oersted field in more detail, we consider first the case of the weak field $\mathbf{H}_{Oe}=0.1 \cdot \mathbf{H}_{inf}$. For small currents we observe again the propagating mode $W_2$ similar to that found in the absence of $\mathbf{H}_{Oe}$. As explained above, the current region where this mode exists is narrower than for $W_1$ ($\Delta I_{W2}$ = 15-17.5 mA for $\kappa = 0.1$, instead of $\Delta I_{W1}$ = 15-20 mA for $\mathbf{H}_{Oe} = 0$). Due to the abovementioned asymmetry of the underlying equilibrium magnetization state caused by the Oersted field (and also due the overlapping of the Oersted field with the homogeneous external field $\mathbf{H}_0$ = 30 Oe used in all our simulations presented here), the group velocity of a spin wave significantly depends on the propagation direction. Hence magnetization pattern for the $W_2$-mode becomes circularly asymmetric (see gray-scale maps in Fig. 6).

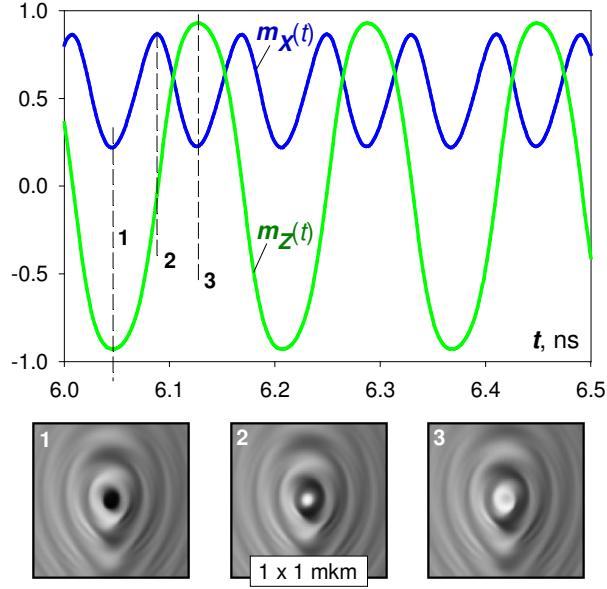

Fig. 6. The same as in Fig. 2 for the propagating mode in the single layer system for $\mathbf{H}_{Oe} = 0.1\mathbf{H}_{inf}$.

By higher currents the propagating wave mode vanishes, and the localized modes appear. In presence of the Oersted field these modes differ qualitatively from localized modes when $\mathbf{H}_{Oe}$ is neglected. In particular, it is instructive to compare the localized mode $L_2$ arising at the high current $I = 48$ mA at $\mathbf{H}_{Oe} = 0$ (see Fig. 4) with the localized mode $L_3$ appearing for the same current, but for $\mathbf{H}_{Oe}=0.1\cdot\mathbf{H}_{inf}$ (shown in Fig. 7). From gray-scale maps and arrow plots shown in Fig. 7 it is clear, that in contrast to the quadrupole V-AV mode in the absence of $\mathbf{H}_{Oe}$, the magnetization configuration of the $L_3$-mode at higher currents even in presence of a relatively weak Oersted field ($\kappa = 0.1$) contains most of the time only a single V-AV pair. This fact can be again attributed to the strong circular asymmetry of the magnetization state in presence of $\mathbf{H}_{Oe}$, so that the formation of a highly symmetric V-AV quadrupole becomes impossible.

Magnetization dynamics of this new $L_3$ mode is also completely different. Due to the same reason - circularly asymmetric equilibrium magnetization configuration - the V-AV pair can not rotate free around its center even when vortex and antivortex have opposite polarities. Magnetization dynamics of the $L_3$-mode is thus governed by the *oscillation of the vortex position*, whereby the antivortex remains nearly immobile. During these oscillations the vortex generates a 2$^{nd}$ V-AV pair with a very small V-AV distance and the same polarities, which are, however, *opposite* to the polarity of the *initial* vortex. The antivortex from this new small V-AV pair annihilates with the initial vortex (irradiating a burst of spin waves), thus leaving a single vortex with the polarity opposite to the initial vortex, so that effectively the polarity of the vortex in the initial large V-AV pair is reversed. Then this new vortex starts to move in the opposite direction, generates a new small V-AV pair and the process in repeated. A very similar process was reported as a mechanism to change the vortex polarity during a motion of a single vortex in [37].

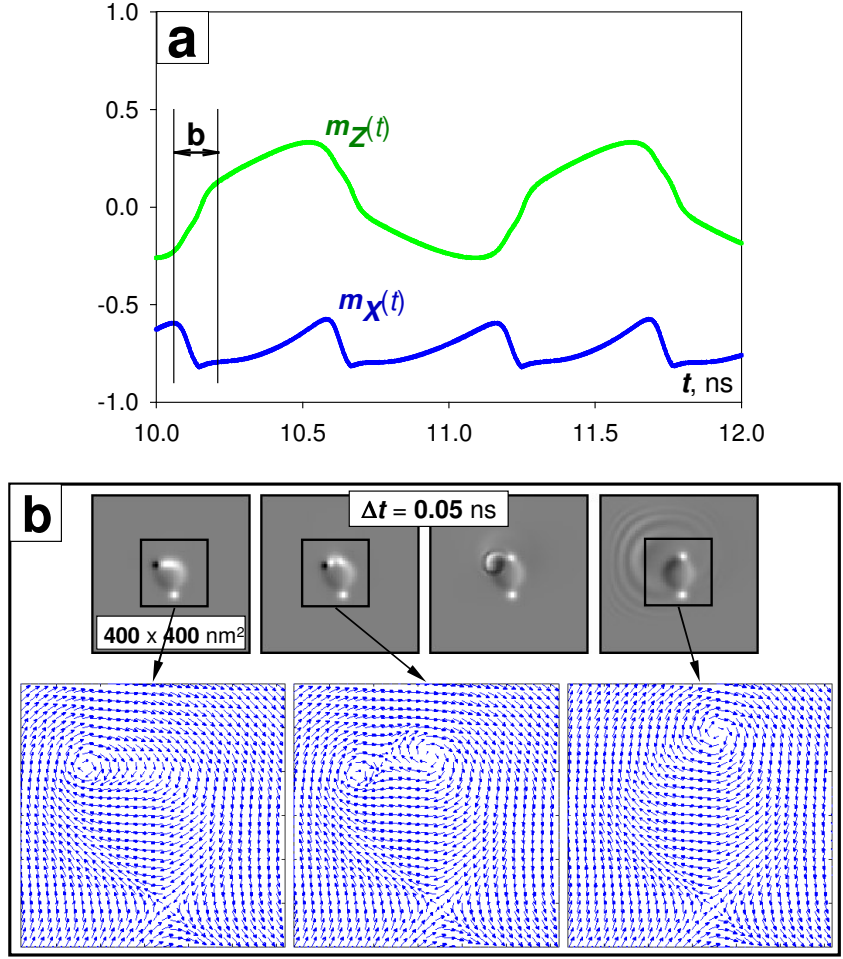

Fig. 7. (a) Magnetization time dependencies for the high-current localized mode $L_3$ in the single layer system for $\mathbf{H}_{Oe} = 0.1\mathbf{H}_{inf}$. (b) Oscillations of the main V-AV pair, accompanied by the creation and annihilation of a satellite V-AV pair (snapshots of the out-of-plane magnetization component $m_y$ and in-plane arrow plots).

When the Oersted field coefficient $\kappa$ is increased further, the propagating mode region contracts even more (right panel in Fig. 5 with $H_{Oe}=0.5\cdot H_{inf}$). Increasing the current for this $\kappa$ value, we still observe after a *W*-mode a relatively narrow region of irregular dynamics, after which a transition to a localized mode $L_4$ similar to that shown in Fig. 7 takes place. Due to the larger Oersted field and stronger deformation of the magnetization configuration the (also immobile) antivortex of this $L_4$-mode is located farther from the point contact center than for the $L_3$-mode at the same current (Fig. 8b). Magnetization dynamics of the $L_4$-mode is governed by the same process of the vortex oscillation accompanied by the creation-annihilation of a small V-AV pair as for the $L_3$-mode. An important point is that the large Oersted field leads for this case to a strong (and approximately linear) increase of the mode frequency with the current strength, because the oscillation frequency of the vortex in the potential well created by the Oersted field near the point contact center increases with the current strength.

Concluding this subsection, we would like to emphasize the following important point: by changing the electric setup (design of the point contact wiring) the strength of the Oersted field can be made different for *the same current strength* - at least up to some extent. As it can be seen from our simulations, such changing of the Oersted field strength can be used to control the dominating oscillation mode of the point contact device.

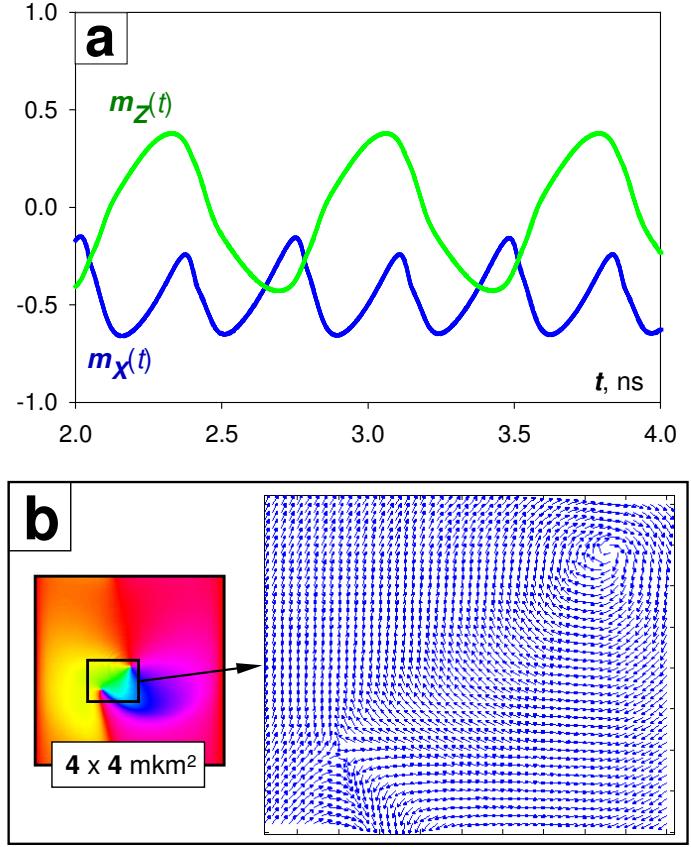

Fig. 8. (a) Magnetization time dependencies for the high-current localized mode $L_4$ in the single layer system for the large Oersted field $\mathbf{H}_{Oe} = 0.5\mathbf{H}_{inf}$; (b) Image of the in-plane magnetization orientation for the whole simulation area and the enlarged in-plane arrow plot of the V-AV pair.

### C. Magnetization dynamics for the CoFe/Cu/Py trilayer system: Influence of the ‚hard' magnetic layer

The next important question for the SPC-induced magnetization dynamics is the influence of the 'fixed' magnetic layer on magnetization oscillations of the 'free' layer. Up to our knowledge, this influence was not studied yet for the point contact setup in the extended thin film geometry, so the results presented below are especially interesting. We remind that the 'fixed' layer parameters have been chosen to imitate the $Co_{90}Fe_{10}$ underlayer used in the system studied experimentally in [14]: $M_S = 1500$ G, $A = 2 \cdot 10^{-6}$ erg/cm, layer thickness $h_{fix} = 20$ nm and the spacer thickness (distance between free and fixed magnetic layers) $h_{Cu} = 4$ nm. Magnetization dynamics of the fixed layer induced by the magnetodipolar interaction between the free and fixed layers was fully taken into account. Further, we assumed that the fixed layer thickness is large enough to neglect the spin torque effect on this layer.

In principle, one should also keep in mind that the magnetocrystalline anisotropy of $Co_{90}Fe_{10}$ is not small (cubic anisotropy with $K_{cub} = -5.6 \cdot 10^5$ erg/cm$^3$) and hence, generally speaking, can not be neglected. However, the influence of this anisotropy for a polycrystalline $Co_{90}Fe_{10}$ material as used in [14] is partially 'averaged out' [38] due to the very small grain size (~10-20 nm). Both for this reason and lacking the exact knowledge about the grain size and texture of magnetic layers studied experimentally in [14], we present here only the results where the magnetocrystalline anisotropy of the fixed layer is neglected. Our preliminary studies show that when this random anisotropy is taken into account, a substantial dependence of the results on the grain size and film texture is observed, so that one should possess a quantitative information about these parameters in order to make a meaningful quantitative comparison of simulated and experimental data.

In order to study solely the effect of the magnetodipolar interaction between the fixed and free layers, we have also neglected the influence of the Oersted field in simulations which results are presented in this subsection.

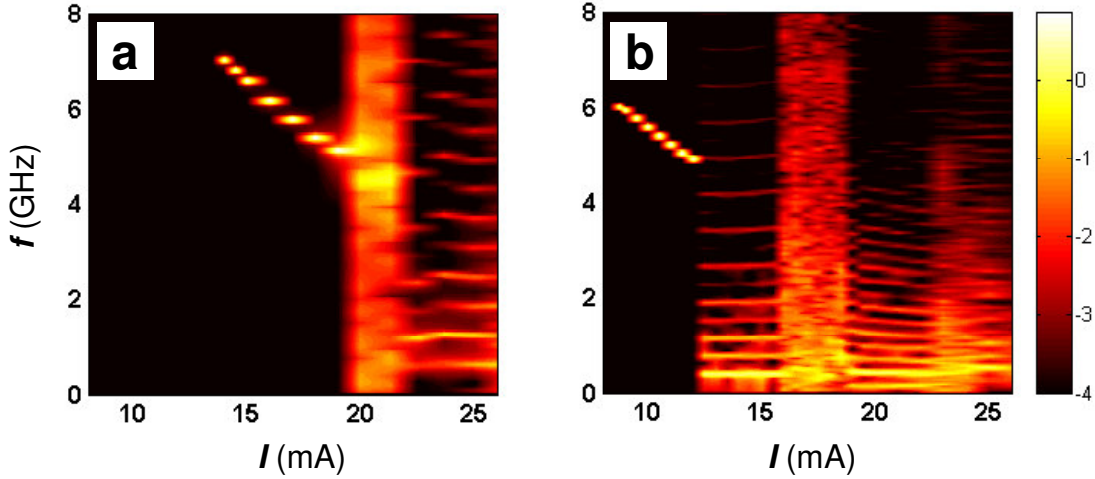

Fig. 9. Comparison of spectral power maps of the $m_z$-component for the single layer (a) and trilayer (b) systems ($\mathbf{H}_{Oe} = 0$). The strong reduction of the current strength for the oscillation onset and for the transition to the localized modes due to the influence of the hard magnetic layer can be clearly seen (the discrete character of the $f(I)$ dependence for $W$-modes is due to the same reasons as explained in Fig. 1).

The first very important result of our simulations is that the magnetodipolar interlayer interaction leads to the drastic reduction of the threshold current $I_{th}$ for the magnetization oscillations onset: from the comparison of oscillation power plots in Fig. 9 it follows, that the presence of the hard magnetic underlayer reduces the threshold current from $I_{th} \approx 13$ mA for the single layer system to $\approx 8$ mA for the trilayer. The current where the transition from the propagating to localized modes occurs, is reduced from $I_{loc} \approx 20$ mA for the single Py layer system to $I_{loc} \approx 12$ mA (i.e., it nearly halves) when the presence of the CoFe underlayer is taken into account.

The most probable qualitative explanation of this effect is the following: when the magnetization of the free layer deviates slightly from its in-plane orientation (under the influence of the spin torque within the point contact area), a stray field is generated. Straightforward geometrical consideration shows that this stray field causes the deviation of the fixed layer magnetization in the direction opposite to that of the free layer. This, in turn, results in the stronger deviation of the free layer magnetization due to the influence of the fixed layer stray field, thus leading to the positive feedback between the magnetization dynamics of the free and fixed layer. Such a positive feedback leads to the decrease of the threshold current for the oscillation onset. An additional decrease of the transition current from the propagating to the localized modes can be explained by the fact that magnetodipolar interlayer interaction field is strongly inhomogeneous, thus favoring the appearance of the localized modes. From the quantitative point of view, however, such a large decrease of the threshold current due to the magnetodipolar interlayer interaction is surprising.

Now we proceed to the discussion of the hard magnetic layer influence on various oscillation modes. As it can be seen from Fig. 9b, for the trilayer system the first appearing mode (for currents slightly higher than $I_{th} \approx 8$ mA) is also the propagating one. The pattern of the spatial wave propagation for this mode is asymmetric (see Fig. 10), what in this case is due to the inhomogeneous magnetodipolar field created by the hard magnetic layer (we remind that the Oersted field is not included into the trilayer simulations in order to isolate the interlayer interaction effect). The frequency at the oscillation onset is substantially lower than for a single-layer system (~ 6 GHz instead of ~ 7 GHz), showing that the interlayer interaction should be also taken into account, if the quantitative comparison between simulated and experimental frequencies is aimed. When the current increases further, the oscillation frequency decreases almost linearly with current, as for a single-layer system.

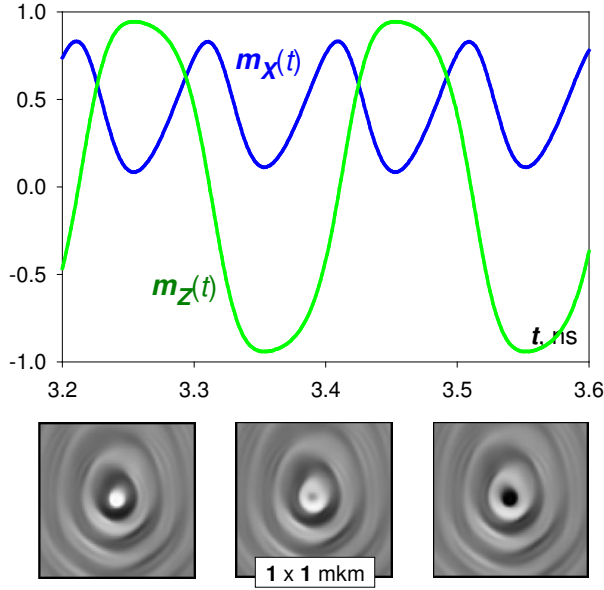

Fig. 10. The same as in Fig. 2 for the propagating (Slonczewski) mode in the trilayer system for $\mathbf{H}_{Oe} = 0$.

At the current value $I_{th}^{loc} \approx 12.3$ mA we find a transition to the first localized mode of our trilayer ($L_5$-mode) with a very low (for the SPC-induced magnetization dynamics) oscillation frequency ~ 380 MHz. Similar to the first localized mode $L_1$ of a single-layer system, the main dynamic process for this mode is the rotation of a V-AV pair with opposite polarities of the vortex and antivortex (Fig. 11b). During this rotation the V-AV distance slowly changes due to the interaction with the hard magnetic layer and presence of a small constant external field. However, in contrast to the $L_1$-mode for a single layer case, here we observe during the rotation of this main V-AV pair the creation of a small (satellite) V-AV pair with *opposite* vortex and antivortex polarities. For this reason this small pair does not propagate (like the satellite pair for the $L_1$-mode), but its antivortex immediately annihilates with the vortex of the main pair, emitting a burst of spin waves (Fig. 11c), similar to localized modes found in the presence of the Oersted field (see Fig. 7b and 8). This very fast creation-annihilation process manifests itself in a small cusp on the time dependence of the $m_z$-projection and a large peak on $m_x(t)$ (Fig. 11a). Due to the extremely anharmonic time dependencies of both in-plane magnetization projections the oscillation power spectrum of this mode contains very strong higher harmonics clearly visible in Fig. 9b.

For the $L_5$-mode it is also possible to compare the rotation frequency of the V-AV pair with that calculated from the theory of Komineas [34]. The typical V-AV distance deduced from simulations for this mode is $d \approx 100$ nm, what results in the analytically calculated (see Eq. (4)) frequency $f^{an} \approx 0.36$ GHz. Simulated frequency for this case is $f^{sim} \approx 0.38$ GHz. Again, we obtain a very good agreement between theory and simulations, although the theory includes neither the magnetodipolar interlayer interaction, nor the process of the V-AV creation-annihilation. As mentioned above, the latter process probably has almost no effect on the rotation frequency of the main V-AV pair, because (compare time scales in Fig. 11b and 11c) the satellite V-AV creation-annihilation happens very fast compared to the rotation period of the main V-AV pair. However, it is not clear why the effect of the hard layer stray field on the rotation of the main V-AV pair is apparently also rather small. We would also like to mention, that in order to enable a more meaningful comparison, one should use not the 'typical' V-AV distance, but the inverse square of this distance $\langle 1/d^2 \rangle$, averaged over the rotation period.

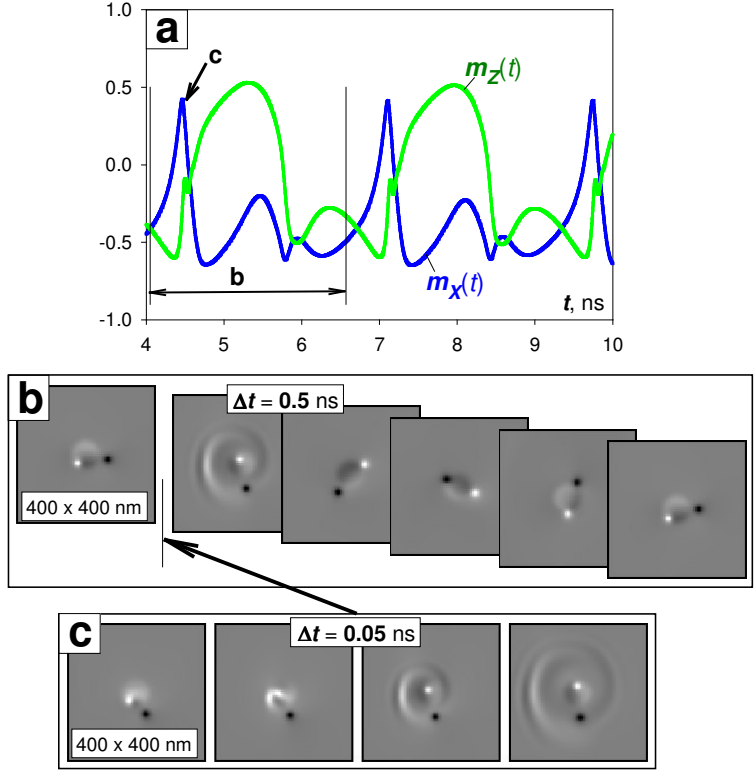

Fig. 11. Magnetization time dependencies (a) and snapshots of the magnetization configurations ((b) and (c)) for low-current localized mode $L_5$ in the trilayer system ($\mathbf{H}_{Oe} = 0$). Panel (b) shows the rotation of the main vortex-antivortex pair (images of the out-of-plane magnetization component) and panel (c) illustrates the creation and annihilation of a satellite V-AV pair accompanied by the emitting of a spin wave burst. Time intervals corresponding to the image rows of (b) and (c) are marked with vertical lines on the plot in the panel (a).

When the current increases above ≈ 16 mA, magnetization dynamics becomes less regular, which manifests itself in a quasicontinuous power spectrum up to the current value ≈ 19 *mA*, where the next regular dynamic mode ($L_6$) appears. The major spectral peak of this mode has approximately the same frequency $f \approx 0.38$ GHz as for $L_5$, but the analysis of the magnetization configurations reveals, that the complete period of $m_z$-oscillations $T_z \approx 7.8$ ns corresponds to an ever lower frequency $f_z \approx 0.13$ GHz; corresponding relatively weak spectral band can be also seen in Fig. 9b.

This high-current mode is the most complicated among regular modes studied here and combines all the processes analyzed above (see Fig. 12). Its formation starts from the nearly homogeneous magnetization deviation under the point contact area, which evolves very rapidly into a V-AV pair with the same polarities of the vortex and antivortex (Fig. 12, image row A). This process is up to some extent similar to the formation of the high-current mode $L_2$ for a single layer system (Fig. 4), but in the single layer case *two* V-AV pairs were formed.

During the next stage (row B in the same figure) the V-AV distance in this pair increases, and the pair orientation is slightly changed, what is possible due to the magnetodipolar field of the hard layer (as mentioned above, the V-AV pair with the same V-AV polarities could move only translationally in the absence of external fields). At the third stage (row C in Fig. 12) the smaller V-AV pair is formed near the vortex of the main pair. The polarities of vortex and antivortex in this satellite pair are the same, but *opposite* relative to the V-AV polarities of the main pair. For this reason the antivortex of the new satellite pair annihilates very fast with the vortex of the main pair, emitting a burst of spin waves, as explained above.

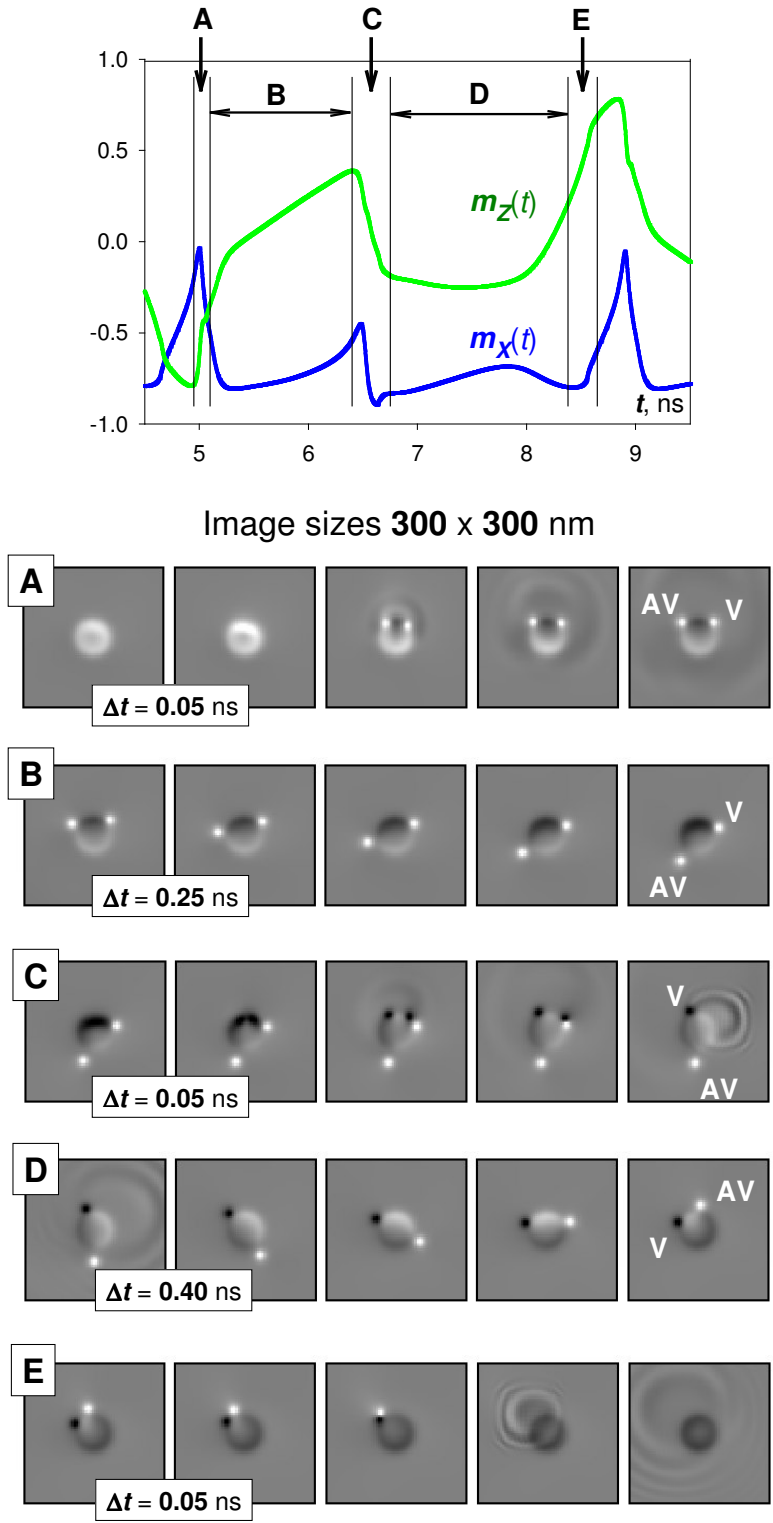

Fig. 12. The same as in Fig. 11 for the high-current localized mode $L_5$ in the trilayer system ($\mathbf{H}_{Oe} = 0$). Panels A-E show various dynamic processes constituting this very complicated mode; see text for details. Corresponding time intervals are marked on the plot in the panel (a).

As a result of this annihilation, the V-AV pair with nearly the same V-AV distance as at the very beginning of the process, but *opposite* polarities of vortex and antivortex, is left. Due to opposite V-AV polarities, this pair starts to rotate (row D in Fig. 12), and the distance between vortex and antivortex continuously decreases (probably also due to the influence of the underlayer stray field), until they annihilate (row E). The nearly homogeneous magnetization deviation under the point contact area left after this annihilation is directed *opposite* to the initial deviation at the start of the process. This means that the images displayed in the rows A-E in Fig. 12 correspond to a half of the complete magnetization dynamic period of this mode.

We would also like to emphasize, that for all localized modes found in the system when the Oersted field was neglected ($L_1$, $L_2$, $L_5$ and $L_6$), the mode frequency was nearly independent on the current strength, although some of these modes existed in a quite large current region (e.g., modes $L_5$ and $L_6$). The most probable explanation of this interesting feature is that the mode frequency is determined by the rotation frequency of the V-AV pair(s) constituting the mode. This rotation frequency, in turn, is governed by the V-AV separation within the pair, which in our system is defined mainly by the point contact diameter flooded by the current (in the absence of the Oersted field) and additionally - by the stray field of the underlayer (for the trilayer system). For this reason the mode frequency does not change noticeably with the current strength. The increasing amount of energy pumped into the system when the current strength grows is probably 'consumed' during the process of the creation-annihilation of satellite V-AV pairs discussed above. Indeed, a micromagnetic analysis has shown that the magnetization configuration of these pairs depends on the currents strength.

When the Oersted field is included, and it is large enough, the antivortex becomes immobile, and the mode frequency is determined by the oscillation of the vortex position within the potential well created by the Oersted field. For this reason the mode frequency increases with current, because larger currents create stronger Oersted fields. This frequency increase is especially pronounced for large $\kappa$ (panel for $\kappa = 0.5$ in Fig. 5),

### D. Comparison with experimental results

At present there exist only very few experimental studies (partially supported by numerical simulations) of the SPC-driven microwave oscillations in the point contact geometry, where it is proven - or at least suggested with a high plausibility degree - that this dynamics is governed by the vortex/antivortex motion [14, 24, 25]. Since we have chosen our simulation parameters according to the device characteristics from [14], we shall mainly compare our results with those reported in this paper.

First we note that we have simulated the in-plane field geometry, so that our results should be compared with the experimental data reported in the first part and in Fig.1 of [14]. It also follows from our simulations that for this particular setup, the strong influence of the interlayer interaction on the power spectra of the microwave oscillations can not be neglected (see Sec. III.C above). Hence only simulated data obtained for the complete trilayer system (see Fig. 9) can be used for a meaningful comparison with the experiment.

This comparison shows that our simulation results for the first localized mode of the trilayer (mode $L_5$, see Fig. 11) could reproduce semiquantitatively several important features of the real experiment. First, the current region where this mode is observed experimentally ($\Delta I_{\exp} \approx$ 6 - 12 mA) (see Fig.1 in [14]) is close to the region where our $L_5$ mode is found numerically ($\Delta I_{\text{sim}} \approx$ 12 - 16 mA). The frequency of the experimentally observed microwave oscillations $f_{\exp} \approx$ 100 - 220 MHz [14] is of the same order of magnitude as the simulated frequency $f_{\text{sim}} \approx$ 380 MHz. Next, the weak dependence of the experimentally measured frequency on the current strength corresponds fairly well to our observation that the frequency of the localized modes for a trilayer is nearly current-independent. Finally, the strongly non-sinusoidal character of simulated magnetization oscillations is in accordance with the presence of several higher harmonics in the experiment [14].

The comparison of experiment and simulations could be more meaningful, if a better characterization of the experimentally studied system would be available. The problem is not only that several important geometric parameters (e.g., the point contact diameter) are not known exactly. We have also found, that the threshold current $I_{th}$ for the oscillation onset depends on the average characteristics of the crystallographic structure of the hard layer (grain size and texture), and, what is less evident - that $I_{th}$ is also different for various particular realizations of a random grain structure with the same average parameters. This difference is caused by substantial variations of the equilibrium magnetization configuration of the hard layer for various random grain structure realizations. The oscillation frequency itself and even the type of the first localized mode also depend on the crystallographic structure of the hard layer; it is worth noting here that the fundamental frequency of, e.g., our $L_6$-mode ($f_6 \approx 130$ MHz) lies within the experimentally measured frequency range.

Further, we have observed, that when the Oersted field is taken into account (even strongly weakened with respect to a maximal possible field of an infinitely long nanowire), the type of the first localized mode could be changed. The Oersted field also causes the increase of the oscillation frequency with current similar to the frequency behavior observed experimentally.

There exist, however, important issues, where we observe a qualitative disagreement with experimental findings. The most important one is the presence of several well defined different oscillation modes in the simulated dynamics, whereby experimentally only one oscillation mode was found (this statement is based mainly on the absence of any jumps on the current dependencies of the oscillation frequency and power). The absence of other localized modes at currents higher than the 'switch-off' current for the 1$^{st}$ localized mode in the experiment can be in principle explained as follows: according to [14], when the current is increased, the well-defined spectral peak evolves into "a broader band spectral output at larger current". This observation is consistent with the spectrum evolution found in our simulations, as demonstrated by the transition from a spectrum consisting of sharp spectral lines to a broadband spectrum at $I \approx 16$ mA in Fig. 9b. The next localized mode could be suppressed in a real experiment, because it emerges at much larger currents, when the sample heating and/or spin torque fluctuations due to the high spin current density prevents the formation of a well-defined oscillation mode.

The absence of the propagating mode (*W*-mode in our notation) is more difficult to explain. In principle, one can speculate that the propagating mode vanishes due to the presence of the Oersted field, which reduces the current region $\Delta I_W$ where the *W*-mode exists (see Fig. 5). However, we note that significant reduction of the current interval $\Delta I_W$ requires high Oersted fields - with the magnitude close to that achieved for a very long contact wire. Taking into account, that in real experiments the length (height) of the cylindrical wire forming the point contact is usually of the same order of magnitude as the contact diameter, this explanation is questionable. Still, it can be fully excluded only when one will be able to perform simulations with the Oersted field computed according to the real electrical setup.

To provide another explanation why the *W*-mode is absent in the experiment [14], we would like to remind that the same discrepancy was found for systems with a small point contact diameter in high external fields. Simulations have predicted that for such systems, when the current *increases*, the propagating mode should emerge first [21, 22], and the localized ('bullet') mode should appear at higher currents [20]). However, experimentally only a localized mode was found [12], which was later unambiguously identified as the 'bullet' mode [21, 22, 26]. The probable explanation of this contradiction was based on the theory of Slavin et al. [20], where it was shown that the threshold current $I_{th}^L$ for the 'bullet' mode is smaller than $I_{th}^W$ for the *W*-mode, but the 'bullet' mode has the finite (and not even small) magnetization oscillation amplitude already at its threshold. For these reason the 'bullet' mode was not observed in simulations made by *increasing* current and at $T = 0$. These arguments were supported by numerical simulations, where it was shown that (*i*) when the current *decreases*, the 'bullet' mode still exists for currents smaller than $I_{th}^W$ [21, 26] and (*ii*) the average energy of the localized mode is smaller at the transition current $W \to L$ for the *increasing* current [22] (so that the *W*-mode is actually metastable). Hence it was suggested, that in systems with a small contact diameter in high external fields, the energy required to excite the localized mode with a finite oscillation amplitude is supplied by thermal fluctuations. These fluctuations excite the *L*-mode at its threshold current $I_{th}^L$, which is smaller than $I_{th}^W$ for the *W*-mode, so that the latter did not emerge at all. Whether this explanation is applicable to our case, where the current flooded area is much larger - so that both thermal energy and the energy for a V-AV pair formation are higher - should be a subject of further studies.

Another problem which we have encountered is the reproduction of the experimentally observed hysteretic behavior of magnetization oscillations when the current was first *increased* to the value where the well-defined mode disappeared and then *decreased* to zero [14]. We could not reliably confirm this observation in our simulations, because we have found that the hysteretic behavior simulated numerically depends on the rate with which the current is decreased (this was not the case for systems with smaller point contact diameter in high external fields). Due to the computer time limitations we had to decrease the current strength from its maximum to the value for which we intended to study the magnetization dynamics within $t_{red} \sim 10$ ns. Such interval was not enough to achieve results stable with respect to further increase of $t_{red}$. Experimentally, where the current is reduced within a macroscopic time interval, the corresponding problem obviously does not exist.

Comparison with other reports on the SPC-induced magnetization dynamics in the point contact geometry [24, 25] is possible only from a qualitative point of view, because systems studied in these papers considerably differ from the system investigated by us. The paper of Mistral et al. [25] also deals with the magnetization dynamics for the point-contact injection in an extended thin film sample, but the contact diameter used there is much larger ($d \approx 200$ nm) and the applied field is out-of-plane and relatively strong ($H_{perp} \approx 2000$ G). According to simulations performed in [25] this field leads to a formation of a single vortex already in the absence of any current, i.e., in the equilibrium magnetization configuration. When the spin-polarized *dc*-current is applied to this configuration, the vortex is driven out of the point contact area and its precession around this area governs the magnetization dynamics observed in simulations and (most probably) experimentally. Because the magnetization dynamics is dominated by a single vortex motion, results of [25] can not be directly compared to ours.

In a very recently published paper Finocchio et al [24] have studied magnetization oscillations in a multilayer nanopillar device with the elliptical cross-section and relatively small lateral dimensions 250 x 150 nm$^2$. The current was injected into the nanopillar via a small point contact with the diameter up to $d \approx 40$ nm. Experimentally magnetization oscillations with the frequency $f \approx 0.8$ GHz were observed for zero and small applied fields, whereby the oscillation frequency was found to be nearly current-independent. Supporting numerical simulations [24] have shown that under these experimental conditions the creation and subsequent rotation of a single V-AV pair (analogous to our simplest localized modes discussed in detail above) can take place with a frequency quite close to the values measured experimentally. The simulated rotation frequency of such a pair also was almost independent on the current, in a quailtative agreement with our findings. However, a more detailed comparison of our data with the results from [24] is not really meaningful: the small lateral size of the nanopillar device studied in [24] is important not only by itself (strongly changing, e.g., conditions for the existence of a propagating mode), but also because it leads to the large influence of the shape anisotropy (stray field of the nanopillar borders), which may qualitatively affect the magnetization dynamics of localized modes. Here we would like only to mention that the absence of other dynamic modes and processes of the creation-annihilation of satellite V-AV pairs (found by us for an extended thin film geometry) is most probably due to this small lateral size of the nanopillar studied in [24].

## IV. CONCLUSION

We have presented a detailed study of the magnetization dynamics induced by a spin polarized current injected via a point contact into an extended magnetic multilayer for the case, when the point contact diameter ($D_c$ = 80 nm) is relatively large compared to systems studied previously [12, 13, 20, 21, 22, 23] and the in-plane external field ($H_0$ = 30 Oe) is very small. Under these conditions the system exhibits a rich variety of well-defined oscillation modes, which can be divided into propagating and localized ones.

The frequency of *propagating* modes in the simplest case (when the Oersted field and magnetodipolar interaction between the 'free' and 'fixed' layers are neglected) can be satisfactory described by the Slonczewski theory [18, 31]. However, the agreement between simulated and analytically predicted threshold currents is decisively worse than for the point contact with the much smaller diameter ($D_c$ = 40 nm, see [21, 22, 31]). We assume that this is due to the strong anisotropy of the group velocity of spin waves, emitted out of the point contact area for the low external field and large point contact diameter (both factors lead to relatively low oscillation frequencies). Inclusion of the Oersted field and/or interlayer interaction narrows the current region for the existence of propagating modes and results in an asymmetric wave propagation pattern. This asymmetry is due not only to the influence of the Oersted and/or 'fixed' layer stray fields on the spin wave propagation itself, but also due to the deformation of the equilibrium magnetization configuration of the 'free' layer by these fields.

When the Oersted field is neglected, *localized* modes for the system studied here are governed by the rotational and translational movement of V-AV pairs, in contrast to the small contact diameter case, where the dominating mode was the non-linear 'bullet' [20, 21, 22]. The simulated rotation frequency (for pairs with opposite polarities of V and AV) and translational motion velocity (for pairs with the same polarities of V and AV) for the *steady-state* motion of the V-AV pairs are in a good quantitative agreement with the theory of Komineas et al [34, 35], which employs the scaling arguments familiar from the non-linear dynamics. However, the actual dynamic modes involve much more complicated processes, in particular, the creation/annihilation of additional satellite V-AV pairs (which seem to play an important role for the energy emission out of the point contact area) and creation/annihilation of the V-AV quadrupoles. These processes obviously require further investigation to achieve their deeper understanding.

We have also shown that the inclusion of the Oersted field can lead to qualitative changes of magnetization oscillation modes. In particular, for sufficiently large Oersted fields, the dynamics of V-AV localized modes are dominated by the oscillation of vortex in the potential well created by the Oersted field, whereby the antivortex is nearly immobile. This offers (in principle) a possibility to control the dominating magnetization dynamic mode by adjusting the electric contact setup, which is responsible for the Oersted field strength and configuration.

Finally, we have demonstrated that the magnetodipolar interlayer interaction is qualitatively important for the understanding of the magnetization dynamics in point contact systems at low external fields, leading both to a strong decrease of the threshold current for the oscillation onset and to qualitative changes in the observed magnetization oscillation modes.